\documentclass[%
 reprint,
 amsmath,amssymb,
 aps,
]{revtex4-1}
\usepackage[usenames, dvipsnames]{color}
\usepackage[autostyle]{csquotes}
\usepackage{pgfplots}
\usepackage{tikz}
\usepackage{graphicx}
\usepackage{dcolumn}
\usepackage{bm}

\begin{document}

\preprint{IZTECH-P-04/17}

\title{Induced Affine Inflation}

\author{Hemza Azri}
 \email{hemzaazri@iyte.edu.tr}
\author{Durmu\c{s} Demir}%
 \email{demir@physics.iztech.edu.tr}
\affiliation{%
 Department of Physics, \.{I}zmir Institute of Technology, TR35430
\ \.{I}zmir, Turkey
}%

\date{\today}

\begin{abstract}
Induced gravity, metrical gravity in which gravitational constant arises from vacuum expectation value of a
heavy scalar, is known to suffer from Jordan frame vs. Einstein frame ambiguity, especially 
in inflationary dynamics. Induced  gravity in affine geometry, as we show here, leads to an emergent 
metric and gravity scale, with no Einstein-Jordan ambiguity. While gravity is induced by the vacuum expectation value of the scalar field, nonzero vacuum energy facilitates generation of the metric. Our analysis shows that induced gravity results in a relatively large  tensor-to-scalar ratio in both metrical and affine gravity setups. However, the fact remains that the induced affine gravity provides an ambiguity-free framework. 
\begin{description}
\item[Keywords]
Affine inflation, Induced affine gravity, Induced gravity, Jordan-Einstein frame ambiguity.
\end{description}
\end{abstract}

\pacs{Valid PACS appear here}
\maketitle


\section{Introduction and motivation}
\label{sec:1}
Synthesizing gravitational and field-theoretic dynamics has always been a unifying endeavour. Generating the gravitational constant (the fundamental scale of gravity, $M_{Pl}$) from field-theoretic scales is one such endeavour. Indeed, $M_{Pl}$ has been shown to derive from
the ultraviolet boundary $\Lambda_{\text{UV}}$ of the standard model (SM) as in \cite{sakharov0}, or from the vacuum expectation value of  a non-SM scalar field \cite{zee}. Each option has its motivations. In the present work, we will focus on the second option, given that various phenomena like flavor, baryogenesis, strong CP problem are already modeled heavy scalars. The mechanism is based on a scalar field $\phi$ which directly couples to the spacetime scalar curvature $R\left(g\right)$ to have the action \cite{zee}
\begin{eqnarray}
\label{induced metric action}
S=\int d^{4}x \sqrt{||g||}\left[\frac{1}{2}\xi \phi^{2}R\left(g\right)-\frac{1}{2}
g^{\mu\nu} \nabla_{\mu}\phi\nabla_{\nu}\phi -V(\phi)  \right],\nonumber
\\
\end{eqnarray}
where $\xi$ is dimensionless constant, and $g_{\mu\nu}$ is the metric tensor.  This action is based upon a crucial assumption: There is
no bare gravitational constant to have the Einstein-Hilbert action. If the total energy in this system is minimized at a nonzero field
value $\phi = v$ then, in the vacuum, the fundamental scale of gravity arises spontaneously 
\begin{eqnarray}
\label{newton constant}
M_{Pl}^2 = \xi v^2
\end{eqnarray}
which must have a numerical value $M_{Pl}\approx 2.4\times 10^{18}\ {\rm GeV}$ for this whole mechanism to make physical sense. 
This mechanism is and will be called \textit{Induced Gravity} (IG); it is a theory of gravity based on a scalar-tensor theory 
and it leads to general relativity (GR) in the vacuum albeit with a quantum of the scalar field \cite{zee}.

Though it correctly leads to GR,  the IG is far from providing a complete picture of how the metric tensor itself emerges. This point is important because emergence of gravity starts with a curved metric or curvature. Indeed, classical gravity in its germinal form is a theory of the spacetime metric. It represents the gravitational field as curvature effects on the meter sticks and clocks.  It is this metric 
elasticity  which gives rise to gravity at large distances. To this end, it could be interesting to see if one can generate the metrical elasticity of space. Sakharov's induced gravity \cite{sakharov0} accomplishes this via loops of matter in a curved background.  In the IG based on action (\ref{induced metric action}), however, it is postulated from the scratch to be a Lorentzian manifold so that generation of the Einstein-Hilbert action does not mean induction of {metrical elasticity}. 

A dynamical origin for the metric tensor, through nonzero vacuum energy, has already been proposed and analyzed in the recent work \cite{affine inflation}. There, metric tensor and its equations of motion (gravitational field equations) emerge through a nonzero vacuum energy. Remarkably, nonminimal coupling dynamics in affine gravity is equivalent to a minimal coupling dynamics with a modified potential. It thus turns out that the minimal coupling case must be equivalent to GR after inducing the metric tensor.  

In the present paper, we show that the aforementioned affine gravity can also be induced via the vacuum expectation value of a scalar $\phi$ (as in the action (\ref{induced metric action}) above). To set the stage, we are in an affine spacetime which is endowed with an affine connection, only. What is known are only geodesics, with no notion of angles and lengths. These properties start changing when matter kicks in and, as a result, notion of potential energy crystallizes. Indeed, when a scalar field $\phi$ enters the affine geometry it becomes possible to identify its potential energy. Naturally, vacuum expectation value of the potential energy sets the notion of metric tensor as the energy-momentum tensor of vacuum. Then, the nonzero field configuration that leads to the notion of metric induces the Planck scale through direct coupling between the affine curvature and the scalar field. This framework, which will be called \textit{Induced Affine Gravity} (IAG), will be discussed in Sec. \ref{sec:affine approach}.

Constructing the IAG, we naturally turn to inflationary dynamics where we put special emphasis on induced inflation (in the language of \cite{zee}). Induced affine inflation, as we will call it, will have the Universe undergoing a rapid power-law expansion, starting with small field values ($\phi \ll v$) and gracefully leaving this phase at the field value $\phi=v$. In this setup, the exit is accompanied by a small nonzero cosmological constant (the observationally required value). A remarkable feature of this induced affine inflation is that, in addition to the nearly scale invariant spectrum of perturbations, it predicts a unique spectral index due to the existence of a unique frame set by a unique metric tensor. This feature is an important advantage compared to the induced inflation based on action (\ref{induced metric action}), which suffers from Jordan-Einstein frame ambiguity (see the old and recent works \cite{turner,fakir induced inflation,fakir-frame independent,kaiser,kaiser2,kaiser-frame independent,cerioni,giudice,brans-dicke-planck,karam}). {Nevertheless, here we emphasize that, as in the induced gravity inflation, the observable quantities, namely the spectral index and the tensor-to-scalar ratio, are both sensitive to the nonminimal coupling parameter and can hardly stay in the observational bounds. This of course is not specific to the induced affine inflation; it seems to be a generic feature of the models in which gravity is induced by the vacuum expectation value of a scalar field. The induced affine inflation will be discussed in Sec. \ref{sec:inflationary dynamics}.
     
In section \ref{sec:summary} we conclude.

\section{Induced Affine Inflation}
\label{affine inflation}
\subsection{Induced Gravity: Affine Approach}
\label{sec:affine approach}

Endowed with a symmetric connection $\Gamma^{\lambda}_{\mu\nu}=\Gamma^{\lambda}_{\nu\mu}$ but no metric tensor at all, affine spacetime possesses only one single tensor structure: the curvature tensor. Then, incorporating the scalar field $\phi$, one writes for the invariant action
\begin{eqnarray}
\label{standard affine action}
S \left[\Gamma, \phi \right]= \int d^{4}x \frac{\sqrt{\left| \right| M_{Pl}^2 R_{\mu\nu}\left(\Gamma\right)- \nabla_{\mu}\phi\nabla_{\nu}\phi \left| \right|}}{V(\phi)}
\end{eqnarray}
where the sign $\left| \right|\,\,\left| \right|$ denotes the absolute value of the determinant of the quantities inside.
 
Here, general coordinate transformations of the volume element and the determinant compensate each other to 
lead to an invariant integral. The determinant in the integrand involves a specific combination of the Ricci tensor $R_{\mu\nu}\left(\Gamma\right)$ and the scalar field  kinetic structure $\nabla_{\mu}\phi\nabla_{\nu}\phi$. Its specific nature does not cause any loss of generality. The reason is that general structure of the form $M^2 R_{\mu\nu}\left(\Gamma\right) - c^{2} \nabla_{\mu}\phi\nabla_{\nu}\phi$, with $M$ a mass parameter and $c$ a dimensionless constant, reduces to that in the action (\ref{standard affine action}) after rescaling with $c^{2}$, including 
$c^{-4}$ into a redefinition of the potential energy $V(\phi)$, and finally identifying $M/c$ with the fundamental scale of gravity $M_{Pl}$. Furthermore, the minus sign in front of $c^{2}$ is by convention; it can be reversed by negating $M_{Pl}^2$ everywhere in dynamical equations. These features ensure that the action is general enough to be used for further analysis. It was already analyzed in detail in \cite{affine inflation}. 
 
This action provides a dynamical origin to the metric tensor and it leads to the Einstein equations in GR  with a canonical scalar field $\phi$ ( as studied in \cite{affine inflation}). 

Our goal in this section is to induce the fundamental scale of gravity in the philosophy of the GR counterpart (\ref{induced metric action}), and determine the dynamics of the resulting system. Assuming (as in (\ref{induced metric action})) that there is no bare gravitational constant, the IAG is set by the affine action 
\begin{eqnarray}
\label{induced affine action}
S \left[\Gamma, \phi \right]= \int d^{4}x \frac{\sqrt{\left| \right| \xi \phi^{2} R_{\mu\nu}\left(\Gamma\right)- \nabla_{\mu}\phi \nabla_{\nu} \phi \left| \right|}}{V(\phi)}
\end{eqnarray}
where $\xi$ is a dimensionless parameter.

This action has two peculiarities. Firstly, curvature and scalar field both participate in the formation of the invariant volume. Secondly, equation of motion of $\phi$, as studied in (\ref{standard affine action}), ensures that $V(\phi)$ is the potential energy.
(In the absence of $\phi$ its meaning would be obscure), and finiteness of the action requires that $V\left(\phi \right)\neq 0$. This everywhere-nonzero-potential energy requirement proves important especially in the early Universe where $\phi$ necessitates a nonzero potential to have inflation completed \cite{guth,linde1,albrecht,linde2,higgs inflation,bauer}. 

In what follows, we assume that the potential $V(\phi)$ attains its minimum at some energy scale $v$. This simply suggests a potential energy of the form
\begin{eqnarray}
\label{symmetry breaking potential with cc}
V(\phi)=V_{0}+\frac{\lambda}{4} \left(\phi^{2}-v^{2}\right)^{2}.
\end{eqnarray}
where $\lambda$ is a positive coupling constant. By construction, the potential attains its minimum at $\phi=v$, and this leads to a singular action (\ref{induced affine action}). The vacuum energy, $V_0$, important only at small values of $\phi$, can be set, if needed, to the observed value of the cosmological constant. Its presence ensures that a nonzero cosmological constant exists even at the end of inflation (see Fig \ref{fig:ssb potentials} below). The key point is that this vacuum energy is what ensures the presence of a metric tensor because it possesses a non-singular energy momentum tensor. 

\begin{figure}[b]
\centering
    \includegraphics[width=0.4\textwidth]{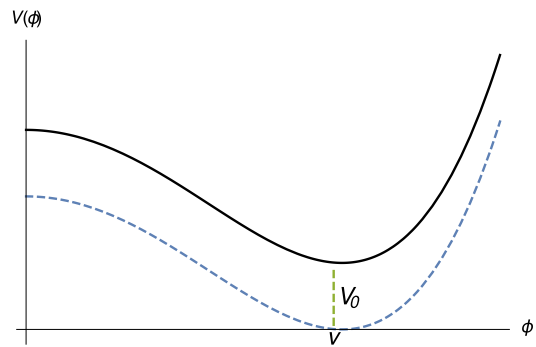}
\caption{Potential energy in (\ref{symmetry breaking potential with cc}) with (solid curve) and without (dashed curve) a nonzero vacuum energy $V_0$.  It is, by definition, trace of the vacuum energy-momentum tensor, and hence, sets therefore the metric tensor.}
\label{fig:ssb potentials}
\end{figure}

Now, the nonminimal coupling term in the action (\ref{induced affine action}) acquires the vacuum expectation value
\begin{eqnarray}
\label{eh}
\xi \left\langle \phi^{2} \right\rangle R_{\mu\nu}(\Gamma),
\end{eqnarray} 
from which follows the fundamental scale of gravity (as defined in (\ref{standard affine action})) 
\begin{eqnarray}
M_{Pl}^2 = \xi v^2
\end{eqnarray}
where $\xi$ and $v$ must have appropriate values to ensure that $M_{Pl}$ takes its correct value. This result, which defines the IAG, proves that gravity can be induced through the vacuum expectation value of the scalar. The difference from the GR, as defined through the action (\ref{induced metric action}), is that:
\begin{enumerate}
\item Vacuum expectation value of the potential energy, $V_0$, introduces the notion of metric tensor (as analyzed in \cite{affine inflation})
\item Vacuum expectation value of the scalar field, $v$, introduces gravity though the coupling term in (\ref{eh}).
\end{enumerate}
These two steps, which should reveal the main difference between the IG and IAG,  show that affine gravity has the potential to accommodate the  emergence of not only the Planck scale (as also happens in the GR) but also the metric tensor. This can be seen more clearly through the equations of motion. Indeed, variation of the action (\ref{induced affine action}) with respect to $\Gamma^{\lambda}_{\mu\nu}$ yields the equation of motion
\begin{eqnarray}
\label{dynamical equation}
\nabla_{\mu} \left\lbrace \xi \phi^{2} \frac{ \sqrt{|| K(\Gamma,\phi)||}}{V(\phi)}
\left( K^{-1} \right)^{\alpha \beta} \right\rbrace = 0, 
\end{eqnarray}
where we have defined for simplicity the following tensor
\begin{eqnarray}
\label{tensor k}
K_{\mu\nu}(\Gamma,\phi)=
\xi \phi^{2} R_{\mu\nu}(\Gamma)-\nabla_{\mu}\phi \nabla_{\nu}\phi.
\end{eqnarray}
It is only after integrating the dynamical equation (\ref{dynamical equation}) that the metrical properties arise. In fact, this equation imposes a specific condition on the connection $\Gamma^{\lambda}_{\mu\nu}$ such that the invertible rank two tensor $g_{\mu\nu}$ which provides a solution to equation (\ref{dynamical equation}) must satisfy
\begin{eqnarray}
\label{density equality}
M^{2}\sqrt{||g||}(g^{-1})^{\mu\nu}=
\xi \phi^{2} \frac{ \sqrt{|| K(\Gamma,\phi)||}}{V(\phi)}
\left( K^{-1} \right)^{\mu \nu}
\end{eqnarray}
and
\begin{eqnarray}
\nabla_{\alpha} g_{\mu\nu}=0,
\end{eqnarray}
where $M$ is an integration constant.

Obviously, the affine connection $\Gamma^{\lambda}_{\mu\nu}$ has now reduced to the Levi-Civita connection $^{g}{\Gamma}^{\lambda}_{\mu\nu}$ of the emergent metric tensor $g_{\mu\nu}$ \cite{affine inflation, azri-eddington, azri-separate, demir-eddington}
\begin{eqnarray}
^{g}{\Gamma}_{\mu\nu}^{\lambda}=
\frac{1}{2}g^{\lambda \sigma}
\left(\partial_{\mu} g_{\sigma\nu}+\partial_{\nu} g_{\mu\sigma}-\partial_{\sigma} g_{\mu\nu} \right).
\end{eqnarray}
It is through $g_{\mu\nu}$ the spacetime geometry acquires metrical structure \textit{a posteriori}. To that end, the gravitational equations which are written in (\ref{density equality}) take the tensorial form 
\begin{eqnarray}
\xi \phi^{2} R_{\mu\nu}(g)-\nabla_{\mu}\phi \nabla_{\nu}\phi
=g_{\mu\nu}V(\phi)\left( \frac{M^{2}}{\xi \phi^{2}}\right) \nonumber \\
\end{eqnarray}   
which can be brought to the standard form through the Einstein tensor 
\begin{eqnarray}
\label{gravitational equations}
\xi \phi^{2} G_{\mu\nu}=&&\nabla_{\mu}\phi \nabla_{\nu}\phi
-\frac{1}{2}g_{\mu\nu}\nabla^{\lambda}\phi \nabla_{\lambda}\phi \nonumber \\
&&-g_{\mu\nu}V(\phi)\left( \frac{M^{2}}{\xi \phi^{2}}\right).  
\end{eqnarray}
These field equations are different from the ones resulting from the action (\ref{induced metric action}) as can be seen from the explicit comparison in \cite{affine inflation}. However, in the vacuum, $\left\langle\phi^{2} \right\rangle=v^{2}$, the affine theory (\ref{induced affine action}) is equivalent to the metric theory (\ref{induced metric action}), and it leads to the Einstein's field equations with cosmological constant if
\begin{eqnarray}
\label{planck mass}
M=\sqrt{\xi} v= M_{Pl}.
\end{eqnarray}
This can be seen from the expression (\ref{density equality}) where the vacuum energy $V(v)=V_{0}$ plays the pivotal role in generating the metric tensor, and it guarantees its emergence. The last step of inducing the metric tensor from vacuum completes the mechanism of inducing gravity. From now on, we assume that all possible contributions to the vacuum energy are incorporated in $V_{0}$, and that they lead to the observed cosmological constant. This means that \cite{affine inflation}
\begin{eqnarray}
V_{0} \sim m_{\nu}^{4},
\end{eqnarray} 
where $m_{\nu}$ is the Neutrino mass.

In conclusion, unlike the metric induced gravity (\ref{induced metric action}) where the metric structure is postulated \textit{a priori}, gravity as a metric elasticity of space is induced in a simple affine space from the affine connection and scalar fields. This emergence not only includes the gravitational constant but also the metric tensor. The IAG stands therefore more extensive than the IG.  

\subsection{Inflationary Dynamics}
\label{sec:inflationary dynamics}

In this section we will analyze inflationary dynamics within the IAG model we constructed above. Now, as can be derived from the action (\ref{induced affine action}), the scalar field obeys the equation of motion 
\begin{eqnarray}
\label{equation of motion of phi}
\Box \phi-V^{\prime}\left(\phi\right)+\xi\phi R\left(g\right)+\Psi\left(\phi\right)=0,
\end{eqnarray}
where the function $\Psi$ is given by
\begin{eqnarray}
\label{psi}
\Psi\left(\phi\right)=\left(1 -\frac{M^{2}}{\xi \phi^{2}} \right)V^{\prime}\left(\phi\right)-\frac{2}{\phi}\left(\nabla\phi \right)^{2}.
\end{eqnarray}

Below, we assume that the Universe is described by the FRW metric with the scale factor $a\left(t\right)$. Then cosmological dynamics of the inflaton $\phi\left(\vec{\text{x}},t\right)$ is described by 
\begin{eqnarray}
\ddot{\phi}+3H\dot{\phi}-\frac{\dot{\phi}^{2}}{\phi}+\frac{(\vec{\nabla}\phi)^{2}}
{a^{2}\phi}-\frac{\vec{\nabla}^{2}\phi }{a^{2}}  \nonumber \\
=\frac{4M^{2}}{\xi \phi^{3}}V\left(\phi \right)
-\frac{M^{2}}{\xi \phi^{2}}V^{\prime}\left(\phi\right), 
\end{eqnarray}
where
\begin{eqnarray}
H^{2}=\frac{1}{3\xi \phi^{2}}\left(\frac{\dot{\phi}^{2}}{2} +\frac{M^{2}}{\xi \phi^{2}}V\left(\phi\right) \right)
\end{eqnarray}
is the Hubble parameter. Inflation proceeds slowly if the slow-roll conditions
\begin{eqnarray}
\frac{\dot{\phi}}{\phi}\ll H,\,\,\,\,\text{and},\,\,\,
\dot{\phi}^{2} \ll \frac{M^{2}}{\xi \phi^{2}}V\left(\phi\right)
\end{eqnarray}
are satisfied. Under these conditions, the background field evolves as
\begin{eqnarray}
\label{slow roll equation1}
&&\ddot{\phi}+3H\dot{\phi}
\simeq \frac{4M^{2}}{\xi \phi^{3}}V\left(\phi \right)
-\frac{M^{2}}{\xi \phi^{2}}V^{\prime}\left(\phi\right),\\
&&H^{2}\simeq \frac{M^{2}}{3\xi^{2}\phi^{4}} V\left(\phi\right),
\label{slow roll equation2}
\end{eqnarray}
which are solved to yield the classical background 
\begin{eqnarray}
\label{solution phi}
&&\phi^{2}\left(t\right)=\phi_{i}^{2}
\pm 4Mv^{2}\sqrt{\frac{\lambda}{3}}\,t, \\
&&\frac{a\left(t\right)}{a_{i}}=\left(\frac{\phi\left(t\right)}{\phi_{i}} \right)
^{1/4\xi}\exp\left\lbrace \frac{1}{8\xi v^{2}}\left(\phi_{i}^{2}-\phi^{2}\left(t\right) \right)  \right\rbrace,
\label{solution a}
\end{eqnarray}
where $\phi_{i}$ and $a_{i}$ are the initial values.

As in the IG, one may consider two behaviors of these solutions depending on the inflationary scenario at hand:
\begin{enumerate}
\item \textit{Chaotic inflation}:
This regime corresponds to initial values $\phi_{i} \gg v$, and then, at early times the scale factor evolves as a quasi-de sitter
\begin{eqnarray}
a\left(t\right) \propto \exp\left\lbrace \frac{M}{2\xi}\sqrt{\frac{\lambda}{3}} \, t \right\rbrace.
\end{eqnarray} 
\item \textit{Ordinary inflation}:
Here the field starts with values $\phi_{i} \ll v$. In this regime, the scale factor is dominated by a power law expansion of the form
\begin{eqnarray}
\label{scale factor power low}
a\left(t\right) \propto t^{1/8\xi}.
\end{eqnarray} 
\end{enumerate}
Here we will focus mainly on the ordinary inflation. Before indulging in the calculation of the spectral index, it is necessary to first write down the equations governing the quantum fluctuation of the inflaton. Then, expanding $\phi\left(\vec{\text{x}},t\right)$ as  
$\phi\left(\vec{\text{x}},t\right)=\phi \left(t\right)+\delta\phi\left(\vec{\text{x}},t\right)$ where the background field
$\phi \left(t\right)$ is given by (\ref{solution phi}), it is easy to see that the fluctuations obey the equation
\begin{eqnarray}
\label{fluctuation equation}
\ddot{\delta\phi}+3H\dot{\delta\phi}+\frac{k^{2}}{a^{2}}\delta\phi
\simeq \frac{\lambda M^{2}v^{2}}{\xi \phi^{2}}\left(1-\frac{3v^{2}}{\phi^{2}} \right)
\delta\phi,
\end{eqnarray}
where $\vec{k}$ is the momentum component corresponding to $\vec{x}$. 

Power spectrum of the scalar perturbations can be calculated using (\ref{fluctuation equation}). However, this may not be straightforward due to the presence of the term on the right-hand side. Nevertheless, we will be interested in the case where the term $k^{2}/a^{2}\left(t\right)$ dominates the term on right-hand side at the time of the last horizon crossing $t=t_{HC}$. In this case, the equation of the fluctuations (\ref{fluctuation equation}) is approximated by the equation of a massless scalar field fluctuation. In fact, for power law $a\left(t\right)\sim t^{p}$, the spectrum of density perturbation is given by \cite{abbott,lyth,sasaki-power law}
\begin{eqnarray}
\label{spectrum of density perturbation}
\mathcal{P} \propto k^{3-2\nu},\,\,\text{with}\,\, \nu=\frac{3p-1}{2\left(p-1\right)},
\end{eqnarray}
which leads to a scalar spectral index $n_{s}$ of the form
\begin{eqnarray}
\label{definition of spectral index}
n_{s}-1 \equiv \frac{d\ln \mathcal{P}}{d \ln k}=3-2\nu.
\end{eqnarray}
In our case, $p=1/8\xi$, and then
\begin{eqnarray}
\label{spectral index}
n_{s}=1-\frac{16\xi}{1-8\xi}.
\end{eqnarray}
This quantity falls in the observational range of the Planck \cite{planck} for {$\xi < 2 \times 10^{-3}$}.

In Table~\ref{tab:2}, we present the spectral index of the inflationary epoch for quasi power-law expansion in a way contrasting the 
IG (based on the metrical action (\ref{induced metric action})) and the IAG, constructed in this paper. The table shows that the predictions of Einstein frame of metrical gravity are close to those of the affine gravity for small $\xi$ values (consistent with observations \cite{planck}). They are, however, essentially different because of the differences in scalar field dynamics (see the discussions below). Needless to say, the IAG is free from the Jordan-Einstein ambiguity present in the IG. 
\begin{table}[b]
\begin{ruledtabular}
\begin{tabular}{cccc}
\textrm{}&
\textrm{IG-Jordan frame}&
\multicolumn{1}{c}{\textrm{IG-Einstein frame}}&
\textrm{IAG}\\
\colrule
\, & \, & \, & \, \\
Power $p$ & $\frac{1}{4\xi}+\frac{3}{2}$ & $\frac{1}{8\xi}+\frac{5}{4}$ & $\frac{1}{8\xi}$\\
\, & \, & \, & \, \\
Tilt $n_{s}$ & $1-\frac{8\xi}{1+2\xi}$ & $1-\frac{16\xi}{1+2\xi}$ & $1-\frac{16\xi}{1-8\xi}$\\
\, & \, & \, & \, \\
\end{tabular}
\end{ruledtabular}
\caption{\label{tab:2} The expansion power (in the form $a\left(t\right)\propto t^{p}$) and the spectral index $n_{s}$ in the IG (which differs between the Einstein and Jordan frames) and in the IAG (which is unique and free from Jordan-Einstein ambiguity). This table should make it clear that a gravity theory like IAG is essential to have unambiguous description of inflation.}
\end{table}

The Einstein-Jordan ambiguity in metrical gravity can be traced back to the conformal transformations that relate the two
frames. The conformal transformation is nothing but a field redefinition and one expects physics in the Einstein and 
Jordan frames to be identical. This is true only  at the classical level, however. The reason is that quantum fluctuations in
the two frames refer to different metric tensors. In fact, passage from Jordan to Einstein frame means removal of the direct mixing between the inflaton and the curvature scalar (proportional to $\xi$) though mixings due to determinant of the metric tensor continue to
exist. In this sense, getting to Einstein frame involves a certain mixture of the metric and the inflaton in the Jordan 
frame, and dynamics of its fluctuations tend to differ from those in the Jordan frame. Saying differently, there arise
difficulties in getting the same result when fluctuation effects are transformed back to the original frame. To elucidate the problem one notes that inflaton fluctuations contribute to the intrinsic curvature perturbation, which is the basis of the {slow-roll approximation} 
underlying the inflationary regime. It turns out that the curvature perturbations in the Einstein (tilded) and in Jordan (not tilded) frames 
are not identical 
\begin{eqnarray}
\label{intrinsic curvature perturbation}
\tilde{\mathcal{R}} \equiv \frac{\tilde{H}}{\dot{\varphi}}\delta \varphi \neq \frac{H}{\dot{\phi}}\delta \phi.
\end{eqnarray}
This means that violation of conformal invariance for curvature perturbations undoubtedly implies different results in different frames. 
In the literature, there have been varying proposals for overcoming the ambiguity \cite{kaiser-frame independent,fakir-frame independent,
sasaki,sasaki-power law,karam,burns,others}, with no obvious resolution yet.

In this respect, the advantage of the IAG is that it provides a unique {geometric} frame (a unique metric). 
The uniqueness of this frame stems from emergence of the metric from the invariant action (\ref{induced affine action}). 
The inflaton $\phi$ propagates in one and the same frame with metric tensor $g_{\mu\nu}$. In fact, 
action (\ref{induced affine action}) can be transformed to a minimal action (\ref{standard affine action}) 
with a new scalar field $\varphi$ by making only a field redefinition of the form (see the earlier studies in \cite{affine inflation})
\begin{eqnarray}
\label{new field}
d \varphi= \frac{M_{Pl}}{\sqrt{\xi}\phi} \, d\phi \,\,\,\,\,\text{and}
\,\,\,\,\, U\left[\varphi \left(\phi \right) \right]= \frac{M^{4}_{Pl}}{\xi^{2}\phi^{4}}
V\left(\phi\right).
\end{eqnarray}
This can also be checked directly from the gravitational field equations (\ref{gravitational equations}) by 
applying the transformation (\ref{new field}). Indeed, since metric tensor remains the same for both minimally and nonminimally-coupled scalars, predictions of AG are protected from mixings of the scalar and 
tensor perturbations arising from conformal transformation of the metric.  A unique gravitational frame ensures 
therefore invariance of the intrinsic curvature perturbations as well as the uniqueness of the 
spectral index (\ref{definition of spectral index}).

The power law inflation that we have studied here is highly illustrative to demonstrate the impossibility 
to get identical spectral indices in Jordan and Einstein frames in the IG. In fact, different powers 
$p$ that correspond to the expansion of the scale factor $a\left(t\right) \propto t^{p}$ in different 
frames lead to different forms of $\nu$ in (\ref{spectrum of density perturbation}) and then to different 
$\vec{k}$ dependencies of the spectral index $n_{s}$. However, quasi de Sitter solutions which arise 
generally for $p \rightarrow \infty$ yields identical $k$ dependencies in the two frames, and hence, 
lead to the same $n_{s}$. This is precisely the chaotic inflation scenario.      

Recently, it has been shown that in pure affine gravity the spectral index derived from slow-roll conditions 
of the field $\varphi$ coincides with the spectral index calculated at second order in Einstein frame of 
general relativity \cite{affine inflation}. Thus, those earlier results combined with the ones here show that 
Einstein frame may be taken to be the physical in metric theories of 
gravity. Nevertheless, differences from the standard induced gravity are clearly not negligible for 
general couplings. These deviations originate from the nonequivalence of the scalar field dynamics in the
minimal and nonminimal coupling cases \cite{affine inflation}. In fact, in IAG the inflaton dynamics 
is governed by its equation of motion (\ref{equation of motion of phi}) which includes a nontrivial part 
$\Psi(\phi)$. This extra term is not avoidable in the affine dynamics. It leaves its imprints on the power 
law expansion (\ref{solution a}) after solving for the background field. While it maintains its form 
under field redefinition in affine induced gravity, this power law is altered by the conformal transformation 
(Einstein frame) in metric IG, leading to an expansion law different than that of IAG. This 
shows again that it is the nonequivalence of the scalar field dynamics that causes the  differences in 
the predicted results.

Another interesting aspect of this power law inflation concerns the coupling $\xi$. Indeed, large $\xi$ drags the spectral index (\ref{spectral index}) up the observed values. However, in the IAG the power $p$ tends to zero as $\xi$ increases, leaving thus no trace of the expansion (see Fig \ref{fig: power law}).

An important and very useful parameter in every inflationary model is the tensor-to-scalar ratio 
$r$, which measures the power in tensor fluctuations with respect to that in the scalar fluctuations. It
is an indicator of the gravitational waves generation. Production of these primordial gravitational 
waves is not restricted to metric theories but it also holds in the pure affine gravity \cite{affine inflation}. 
It is thus important to shed light on the tensor-to-scalar ratio in the present model. It can be derived from the slow-roll parameter $\epsilon$ in the theory, which reads in terms of the Hubble parameter as
\begin{eqnarray}
\label{epsilon}
\epsilon =-\frac{\dot{H}}{H^{2}}=8\xi,
\end{eqnarray}
where we have used the scale factor given by (\ref{scale factor power low}).
\newline
The slow-roll parameter (\ref{epsilon}) is independent of the field redefinition in (\ref{new field}), and takes therefore the same 
value when calculated in terms of the slowly-rolling field $\varphi$. Thus, the tensor-to-scalar ratio in IAG takes the form
\begin{eqnarray}
\label{r in terms of epsilon}
r =16\epsilon =128\xi
\end{eqnarray}
\begin{figure}[h]
\centering
    \includegraphics[width=0.4\textwidth]{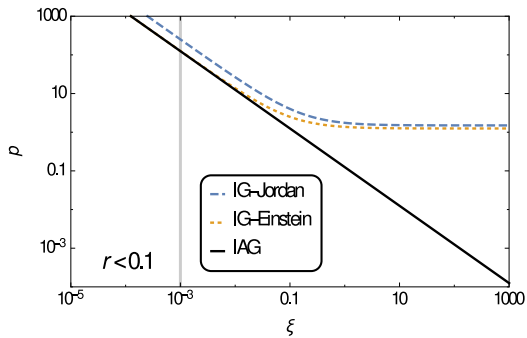}
\caption{The power $p$ as a function of $\xi$, in the IAG and IG. Remarkably, 
for large $\xi$, exit from rapid expansion occurs only in the IAG. The vertical 
line at $\xi=10^{-3}$ corresponds to the observational bound on the tensor-to-scalar ratio.}
\label{fig: power law}
\end{figure}
\begin{figure}[h]
\centering
    \includegraphics[width=0.4\textwidth]{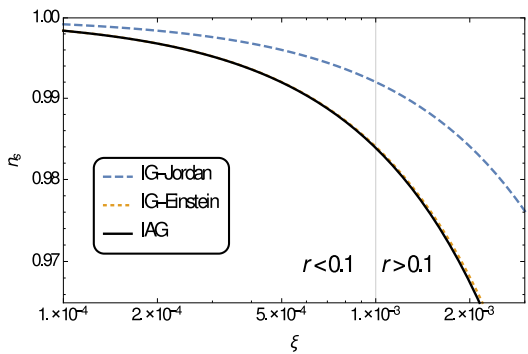}
\caption{The spectral index $n_{s}$ in the IG and IAG.  It is seen that IAG and IG-Einstein stay close to each other. However, the results are generally inequivalent due to the difference of the scalar field dynamics in the two theories. The bound $r<0.12$, corresponding to the vertical line at $\xi=10^{-3}$, pushes the spectral index to larger values. This discrepancy is a feature of induced gravity inflation, may it be IG or IAG.}
\label{fig: first order spectral index}
\end{figure}
Here we must emphasize that the recent data \cite{planck} puts stringent limits on $(n_{s},r)$ which are difficult to satisfy with a single nonminimal coupling $\xi$. In other words, both $n_s$ and $r$ are very sensitive to $\xi$, and thus, the observational bound $r<0.12$ drags $n_s$ outside the observational region. Namely, induced gravity inflation supports mainly large tensor-to-scalar ratio. 
This is not specific to the IAG; it is a generic feature well established in the IG. More specifically, the tensor-to-scalar ratio is given by \cite{burns}
\begin{eqnarray}
r \simeq \frac{128 \xi}{1+6\xi},
\end{eqnarray}
which coincides with (\ref{r in terms of epsilon}) for small $\xi$. The discrepancy between the recent data and the predictions of induced gravity inflation (both IG and AIG) is shown in Fig ~\ref{fig: first order spectral index}, where the spectral index is plotted as a function of the nonminimal coupling $\xi$.
\newline
It is clear that our goal in this work is to construct an ambiguity-free inflationary framework. The discrepancy with the observational data shows that induced gravity inflation (both IG and IAG) may be calling for multi-scalar models. Indeed, in such models nonminimal couplings and  potential landscape can lead to novel configurations bringing agreement with experiment. It is worthy of noting that nonminimally-coupled multiscalars can always be reduced to minimally-coupled scalars in affine gravity \cite{azri-review}, and this is a new feature not found in metrical theories \cite{karamitsos}.

\section{Summary}
\label{sec:summary}

In this work, we have studied induced gravity in metrical and affine theories of gravity. In the first stage, we defined the IG as exists in the literature, and then, we constructed the IAG in affine geometry. We have shown that IAG turns out to be more exhaustive in that it provides a framework in which both metric (through the vacuum energy) and gravity (through the scalar field in vacuum) emerge to lead to a metrical theory. In the second stage, we have studied inflation in the IG and IAG comparatively. We found that IG gives different inflationary parameters in Einstein and Jordan frames. The IG makes no unique prediction that can be contrasted with the observations. The IAG, however,
rests on a unique frame, is thus free from Jordan-Einstein ambiguity, agrees with observations (both spectral index and tensor-to-scalar ratio).  These differences between the IG and IAG are clear enough to motivate the IAG as a viable candidate to study scalar field dynamics as in, for instance, inflation. 

Our findings here can be extended to any other scalar-tensor theory \cite{faraoni-book,extended gravity,maeda-book}. In theories with Jordan-Einstein ambiguity the IAG is expected to reveal interesting physics. The IAG can have a rich phenomenology in both cosmology and astroparticle physics.

\section*{acknowledgments}
This work is supported in part by the T{\"U}B{\.I}TAK grant 115F212.




\bibliography{apssamp}

\end{document}